\documentstyle[referee,epsf]{mn}

\title[Changes in the trajectory of the radio jet in 0735+178?]
{Changes in the trajectory of the radio jet in 0735+178?}

\author[G\'omez et al.] {J. L. G\'omez$^{1}$, J. C. Guirado$^{2}$,
I. Agudo$^{1}$, A. P.  Marscher$^{3}$, A. Alberdi$^{1}$, \newauthor
J. M. Marcaide$^{2}$ and D. C. Gabuzda$^{4,5}$ \\
$^{1}$Instituto de Astrof\'{\i}sica de Andaluc\'{\i}a, CSIC,Apartado
3004, 18080 Granada, Spain\\
$^{2}$Departamento de Astronom\'{\i}a y Astrof\'{\i}sica, Universidad
de Valencia, 46100 Burjassot (Valencia), Spain\\ 
$^{3}$Institute for Astrophysical Research, Boston University, 725
Commonwealth Avenue, Boston, MA 02215, USA\\ 
$^{4}$Joint Institute for VLBI in Europe, Postbus 2, 7990 AA
Dwingeloo, The Netherlands \\ $^{5}$Astro Space Center, P. N. Lebedev
Physical Institute, Leninsky Prospekt 53, 117924, Moscow, Russia\\ }

\date{}

\pubyear{2000}

\begin{document}

\maketitle

\begin{abstract}

We present multi-epoch 8.4 and 43 GHz Very Long Baseline Array images of the
BL~Lac object 0735+178. The images confirm the presence of a twisted jet with
two sharp apparent bends of 90$^{\circ}$ within two milliarcseconds of the
core, resembling a helix in projection. The observed twisted geometry could be
the result of precession of the jet inlet, but is more likely produced by
pressure gradients in the external medium through which the jet propagates.
Quasi-stationary components are observed at the locations of the 90$^{\circ}$
bends, possibly produced by differential Doppler boosting.

Identification of components across epochs, since the earliest VLBI
observations of this source in 1979.2, proves difficult due to the sometimes
large time gaps between observations. One possible identification suggests the
existence of superluminal components following non--ballistic trajectories
with velocities up to $11.6\pm 0.6\,h_{65}^{-1}\,c$. However, in images
obtained after mid-1995, components show a remarkable tendency to cluster near
several jet positions, suggesting a different scenario in which components
have remained nearly stationary in time at least since mid-1995. Comparison
with the earlier published data, covering more than 19 years of observations,
suggests a striking qualitative change in the jet trajectory sometime between
mid-1992 and mid-1995, with the twisted jet structure with stationary
components becoming apparent only at the later epochs. This would require a
re-evaluation of the physical parameters estimated for 0735+178, such as the
observing viewing angle, the plasma bulk Lorentz factor, and those deduced
from these.

\end{abstract}


\section{Introduction}

Radio maps of the BL~Lac object 0735+178 ($z=0.424$, Carswell et
al. \cite{carswell 74}) at milliarcsecond resolution obtained using Very Long
Baseline Interferometry (VLBI) arrays at centimeter wavelengths show a compact
core and a jet of emission extending to the northeast. The first polarimetric
VLBI observations of this source were obtained by Gabuzda, Wardle \& Roberts
\cite{De89} at a wavelength of 6 cm, and revealed a magnetic field
predominantly perpendicular to the jet axis. Multi-epoch VLBI observations of
0735+178 (Cotton et al. 1980; B{\aa}{\aa}th \& Zhang 1991; B{\aa}{\aa}th et
al. 1991; Zhang \& B{\aa}{\aa}th 1991; Gabuzda et al. 1994; G\'omez et
al. 1999) have indicated the existence of superluminal motions with apparent
velocities in the range $\simeq$ 6.5--12.2 $h^{-1}_{65} c$ ($H_{\circ} = 65
h_{65}$ km s$^{-1}$ Mpc$^{-1}$, $q_{\circ}$= 0.5).  Gabuzda et al. \cite{De94}
observed the intersection of a moving and stationary component, during which
there was no evidence for a violent interaction between these features.

\begin{figure*}
\centering
\mbox{\epsfxsize=5.6in\epsfbox[0 0 536 564]{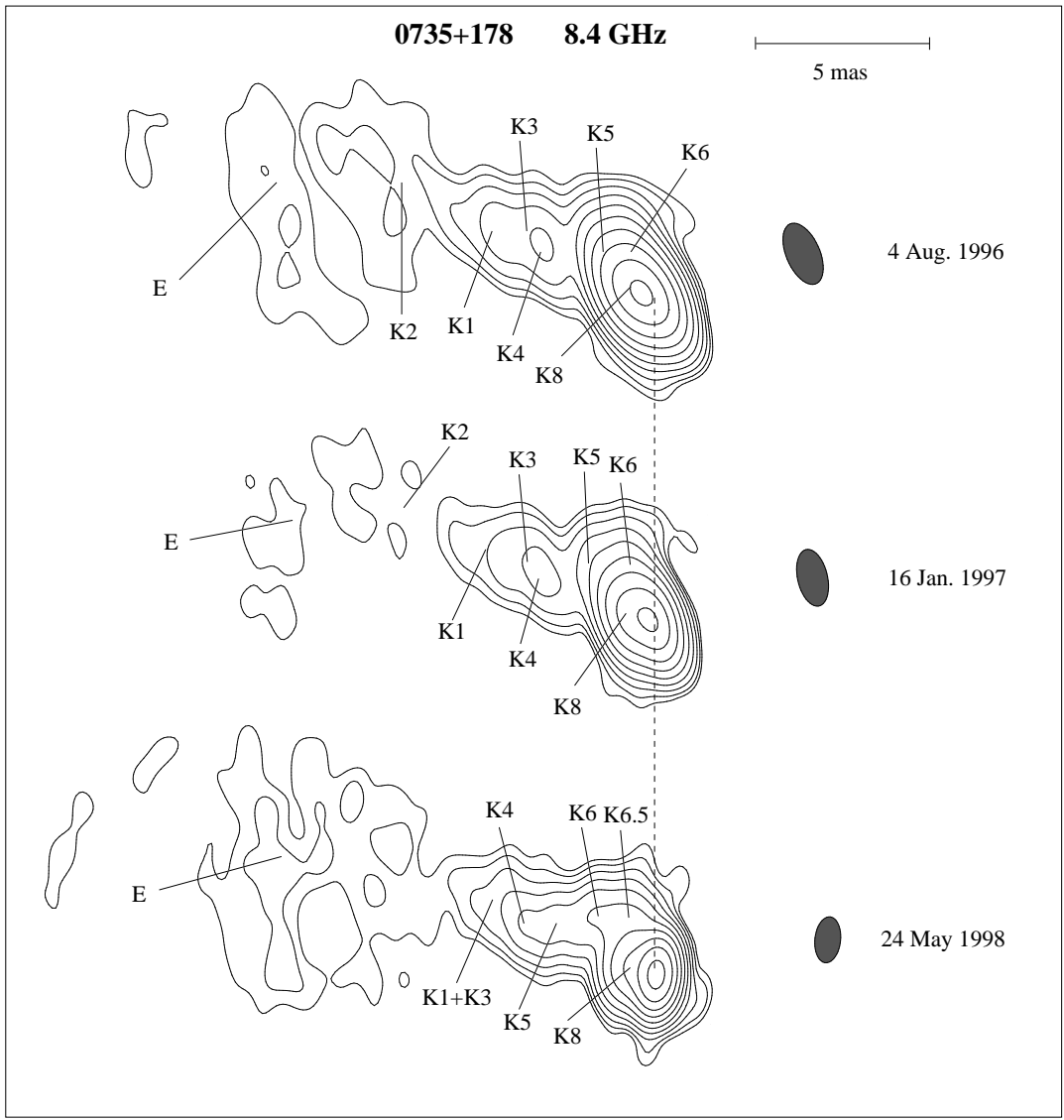}}
\caption{8.4 GHz VLBA images of 0735+178 at epochs (from {\it top to bottom})
4 August 1996, 16 January 1997, and 24 May 1998. Total intensity is plotted as
contours. From top to bottom images: contour levels increment by factors of 2,
starting at 0.2\% (plus 90\%), 0.4\% (plus 90\%), and 0.15\% of the peak
intensity of 1.05, 1.73, and 2.1 Jy/beam; convolving beams (shown as filled
ellipses) are 1.92$\times$0.99, 1.71$\times$0.88, and 1.36$\times$0.76 mas,
with position angles of 23.6$^{\circ}$, 13$^{\circ}$, and -6$^{\circ}$.}
\label{8ghz}
\end{figure*}

\begin{figure*}
\centering
\mbox{\epsfxsize=4.6in\epsfbox[0 0 578 776]{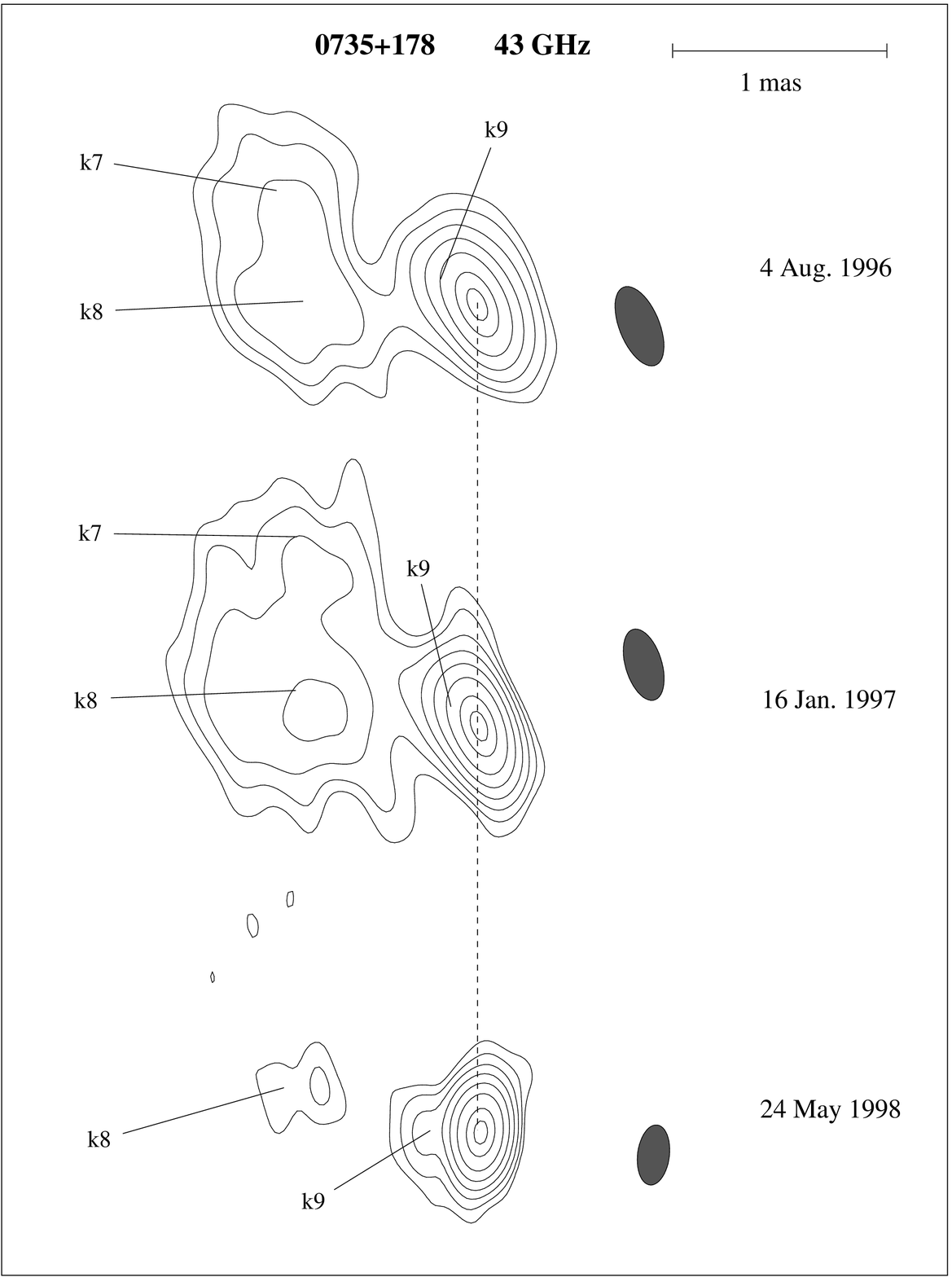}}
\caption{43 GHz VLBA images of 0735+178 at epochs (from {\it top to bottom}) 4
August 1996, 16 January 1997, and 24 May 1998. Total intensity is plotted as
contours. From top to bottom images: contour levels increment by factors of 2,
starting at 2\% (including 90\%), 1\% (plus 90\%), and 1\% (plus 90\%) of the
peak intensity of 0.244, 0.361, and 0.791 Jy/beam; convolving beams (shown as
filled ellipses) are 0.38$\times$0.18, 0.33$\times$0.16, and 0.27$\times$0.14
mas, with position angles of 22.3$^{\circ}$, 17$^{\circ}$, and -7$^{\circ}$.}
\label{43ghz}
\end{figure*}

The first direct evidence of curved structure in the inner jet of 0735+178 was
presented by Kellermann et al. \cite{Ke98}, based on data obtained as part of
a Very Long Baseline Array (VLBA) survey at 15 GHz. Polarimetric VLBA
observations at 22 and 43 GHz by G\'omez et al. \cite{JL99} revealed a twisted
jet with two sharp apparent bends of 90$^{\circ}$ within two milliarcseconds
of the core. The magnetic field appeared to follow smoothly one of the bends
in the jet, interpreted as perhaps produced by a precessing jet nozzle.

In this paper we present new 8.4 and 43 GHz VLBA observations covering three
epochs, aimed at studying structural changes in the jet. The higher resolution
provided by the 43 GHz observations allows mapping of the inner jet components
with an angular resolution of 0.15 mas, through which we can study possible
changes in the direction of the jet nozzle in 0735+178. The higher sensitivity
at 8.4 GHz provides information about the jet motions in the outer regions,
used to test whether components follow ballistic trajectories downstream.

\section{Observations}

  The observations were performed on 4 August 1996, 16 January 1997, and 24
May 1998 using the VLBA at 8.4 and 43 GHz. Left circular polarization data
were recorded at each telescope using 8 channels of 8 MHz bandwidth and 1 bit
sampling. The reduction of the data was performed within the NRAO Astronomical
Image Processing System (AIPS) software in the usual manner (e.g., Lepp\"anen
et al. 1995). Opacity corrections for the 43 GHz observations were introduced
by solving for receiver temperature and zenith opacity at each antenna.

\section{Results}

\begin{table}
\caption[]{8.4 GHz Models for 0735+178.}
\begin{flushleft}
\begin{tabular} {lcccc}
\hline\noalign{\smallskip}
Component$^1$ & $S$ & $r$ & $\theta$ & $FWHM$\\
              & (mJy) & (mas) & ($^{\circ}$) & (mas)\\
\hline\noalign{\smallskip}
\multicolumn{5}{c}{1996.59.}\\
\hline\noalign{\smallskip}
Core &  403$\pm$2 & 0.0 & 0.0 & $\la$0.09\\
K8   &  534$\pm$4 & 0.804$\pm$0.008 & 76.8$\pm$1.5 & 0.45$\pm$0.02\\
K6   &  78$\pm$4 & 1.58$\pm$0.03 & 26.3$\pm$1.3 & 0.58$\pm$0.12\\
K5   &  43$\pm$5 & 2.15$\pm$0.08 & 50$\pm$3 & 0.87$\pm$0.24\\
K4   &  15$\pm$3 & 3.40$\pm$0.21 & 67$\pm$3 & 0.9$\pm$0.4\\
K3   &  23$\pm$3 & 4.27$\pm$0.13 & 65$\pm$2 & 0.8$\pm$0.3\\
K1   &  12$\pm$3 & 5.48$\pm$0.15 & 67$\pm$3 & 0.6$\pm$0.4\\
K2   &  $\la$8 & 7.5$\pm$0.9 & 67$\pm$7 & $\la$3.7\\
E    &  50$\pm$30 & 10.9$\pm$2.4 & 72$\pm$8 & 6.6$\pm$2.3\\
\hline\noalign{\smallskip}
\multicolumn{5}{c}{1997.04.}\\
\hline\noalign{\smallskip}
Core &  542$\pm$11 & 0.0 & 0.0 & $\la$0.3\\
K8   &  533$\pm$15 & 0.836$\pm$0.021 & 72$\pm$3 & 0.60$\pm$0.04\\
K6   &  65$\pm$11 & 1.87$\pm$0.23 & 22$\pm$5 & 0.71$\pm$0.24\\
K5   &  57$\pm$15 & 2.53$\pm$0.24 & 52$\pm$6 & 1.1$\pm$0.5\\
K4   &  19$\pm$11 & 3.7$\pm$0.3 & 69$\pm$8 & $\la$2\\
K3   &  20$\pm$7 & 4.21$\pm$0.23 & 64$\pm$6 & $\la$1.6\\
K1   &  15$\pm$11 & 5.4$\pm$0.4 & 67$\pm$8 & $\la$1.7\\
K2$^2$    & $\sim$10 &$\sim$ 8 & $\sim$66 & $\sim$2.4\\
E   &  $\la$100 & 11$\pm$3 & 74$\pm$15 & 6$\pm$4\\
\hline\noalign{\smallskip}
\multicolumn{5}{c}{1998.39.}\\
\hline\noalign{\smallskip}
Core   &  773$\pm$8 & 0.0 & 0.0 & 0.17$\pm$0.04\\
K8   &  221$\pm$8 & 0.835$\pm$0.020 & 75$\pm$3 & 0.68$\pm$0.03\\
K6.5 (K6) & 60$\pm$11 & 1.79$\pm$0.15 & 23$\pm$3 & 0.76$\pm$0.13\\
K6 (K5) & 71$\pm$7 & 2.43$\pm$0.08 & 46$\pm$3 & 0.91$\pm$0.16\\
K5 (K4) & 45$\pm$7 & 3.35$\pm$0.08 & 64.8$\pm$1.5 & 0.9$\pm$0.3\\
K4 (K3) & 16$\pm$6 & 4.24$\pm$0.3 & 68$\pm$4 & $\la$1.1\\
K1+K3$^3$ (K1) & $\la$30 & 5.2$\pm$0.3 & 65$\pm$3 & 1.2$\pm$0.9\\
E   &  $\la$150 & 11.4$\pm$1.5 & 73$\pm$6 & 6.6$\pm$2.0\\
\noalign{\smallskip}
\hline
\end{tabular}
\end{flushleft}
$^1$Component designations in parantheses correspond to the scenario in
which all components are essentially stationary.\\
$^2$Very extended component for which only estimations of the fitted
parameters could be obtained.\\ $^3$ In the moving-component scenario,
K1 and K3 are merged in a single component with flux density approximately 
that of the sum of both components for epoch 1997.04.
\label{t8ghz}
\end{table}

\begin{table}
\caption[]{43 GHz Models for 0735+178. New model fittings for previous epochs
1996.86 and 1996.98 are also included.}
\begin{flushleft}
\begin{tabular} {lcccc}
\hline\noalign{\smallskip}
& $S$ & $r$ & $\theta$ & $FWHM$\\
Comp. & (mJy) & (mas) & ($^{\circ}$) & (mas)\\
\hline\noalign{\smallskip}
\multicolumn{5}{c}{1996.59.}\\
\hline\noalign{\smallskip}
Core\dotfill & 490$\pm$40 & 0.0 & 0.0 & 0.08$\pm$0.04\\
k9\dotfill & 100$\pm$40 & 0.19$\pm$0.04 & 60$\pm$10 & 0.16$\pm$0.07\\
k8\dotfill & 270$\pm$60 & 0.78$\pm$0.05 & 89$\pm$3 & 0.41$\pm$0.09\\
k7\dotfill & 140$\pm$80 & 1.05$\pm$0.17 & 60$\pm$7 & 0.4$\pm$0.3\\
\hline\noalign{\smallskip}
\multicolumn{5}{c}{1996.86.}\\
\hline\noalign{\smallskip}
Core\dotfill & 249$\pm$4 & 0.0 & 0.0 & $\la$0.05\\
k9\dotfill & 94$\pm$6 & 0.11$\pm$0.02 & 81$\pm$15 & 0.11$\pm$0.04\\
k8\dotfill & 229$\pm$6 & 0.85$\pm$0.02 & 82$\pm$1 & 0.49$\pm$0.03\\
k7\dotfill & 45$\pm$6 & 1.22$\pm$0.09& 54$\pm$4 & 0.30$\pm$0.18\\
k6\dotfill & 48$\pm$17 & 1.65$\pm$0.18 & 24$\pm$7 & 0.94$\pm$0.33\\
\hline\noalign{\smallskip}
\multicolumn{5}{c}{1996.98.}\\
\hline\noalign{\smallskip}
Core\dotfill & 370$\pm$5 & 0.0 & 0.0 & 0.11$\pm$0.03\\
k8\dotfill & 222$\pm$8 & 0.80$\pm$0.04 & 78$\pm$2 & 0.70$\pm$0.06\\
k6$^1$\dotfill & 42$\pm$10 & 1.49$\pm$0.13 & 34$\pm$3 & 0.4$\pm$0.3\\
k3\dotfill & 14$\pm$13 & 3.9$\pm$0.3 & 62$\pm$9 & $\la$0.8\\
\hline\noalign{\smallskip}
\multicolumn{5}{c}{1997.04.}\\
\hline\noalign{\smallskip}
Core\dotfill & 327$\pm$12 & 0.0 & 0.0 & $\la$0.05\\
k9\dotfill & 96$\pm$19 & 0.158$\pm$0.019 & 59$\pm$4 & 0.13$\pm$0.04\\
k8\dotfill & 260$\pm$50 & 0.87$\pm$0.06 & 80$\pm$4 & 0.65$\pm$0.06\\
k7\dotfill & 47$\pm$22 & 1.19$\pm$0.13 & 45$\pm$6 & 0.4$\pm$0.4\\
\hline\noalign{\smallskip}
\multicolumn{5}{c}{1998.39.}\\
\hline\noalign{\smallskip}
Core\dotfill & 536$\pm$8 & 0.0 & 0.0 & 0.059$\pm$0.015\\
k9\dotfill & 73$\pm$19 & 0.189$\pm$0.010 & 80$\pm$20 & 0.26$\pm$0.14\\
k8\dotfill & 60$\pm$40 & 0.89$\pm$0.13 & 78$\pm$10 & 0.5$\pm$0.3\\
\noalign{\smallskip}
\hline
\end{tabular}
\end{flushleft}
$^1$Uncertain identification of this component with K6 at the same epoch.
\label{t43ghz}
\end{table}

Figures \ref{8ghz} and \ref{43ghz} show the VLBA total intensity images of
0735+178 at 8.4 and 43 GHz, respectively. In order to obtain a better
characterization of the motions in the jet, the uv-data were fitted with
circular Gaussian components using the Difmap software package (Shepherd
1997). Tables \ref{t8ghz} and \ref{t43ghz} summarize the physical parameters
obtained for 0735+178 at 8.4 and 43 GHz, respectively.  The process of model
fitting is not deterministic, and often more than one model that fit the data
roughly equally well can be obtained for a given data set. Knowledge of the
source structure at different observing frequencies and different epochs can
help to distinguish among different possible model fits. We have adopted these
criteria, selecting the model fits that not only fit best to the source
structure for a given epoch and frequency, but are also in best agreement with
the known source evolution and structure at different frequencies.

In light of the new observations presented here, we have performed new model
fits of the data presented by G\'omez et al. \cite{JL99}. The new model fits
are tabulated in Table \ref{t22ghz} for the 22 GHz data obtained in 1996.86
and 1996.98, and Table \ref{t43ghz} for the corresponding 43 GHz data. These
revised model fits result in a better identification of components across
epochs, and we have therefore adopted them as the most plausible models for
0735+178. Note, however, that none of the conclusions presented by G\'omez et
al. \cite{JL99} are affected by our adoption of the new model components. The
table columns give the total flux density, separation and structural position
angle relative to the core component, as well as the FWHM angular size of each
component for the three epochs. Components are labeled using uppercase letters
for 8.4 GHz following the notation of Gabuzda et al. \cite{De94}, and are
marked in Figs. \ref{8ghz} and \ref{43ghz}. Estimates of the model fitting
errors were obtained by using the addendum program for DIFMAP called DIFWRAP,
developed by Jim Lovell. The errors were determined by introducing small
changes in the fitted parameters of Tables \ref{t8ghz}, \ref{t22ghz}, and
\ref{t43ghz} and analyzing the variations in the residual maps and $\chi^2$ of
the fit (see http://www.vsop.isas.ac.jp/survey/difwrap/ for more details). In
order to take into account the fact that the fitted parameters are
interrelated, we allowed for simultaneous variations of all four parameters
for each component when determining the final errors, which results in
somewhat larger but more conservative and perhaps more realistic values.

\begin{table}
\caption[]{22 GHz Models for 0735+178 for previous observations in 1996.86 and
1996.98 (G\'omez et al. 1999).}
\begin{flushleft}
\begin{tabular} {lcccc}
\hline\noalign{\smallskip}
& $S$ & $r$ & $\theta$ & $FWHM$\\
Comp. & (mJy) & (mas) & ($^{\circ}$) & (mas)\\
\hline\noalign{\smallskip}
\multicolumn{5}{c}{1996.86.}\\
\hline\noalign{\smallskip}
Core\dotfill & 292$\pm$8 & 0.0 & 0.0 & $\la$0.01\\
K9\dotfill & 75$\pm$6 & 0.18$\pm$0.02 & 95$\pm$13 & 0.20$\pm$0.06\\
K8\dotfill & 192$\pm$6 & 0.86$\pm$0.02 & 85$\pm$1 & 0.51$\pm$0.03\\
K7\dotfill & 189$\pm$10 & 0.93$\pm$0.03 & 61$\pm$3 & 0.76$\pm$0.05\\
K6\dotfill & 30$\pm$4 & 1.66$\pm$0.14 & 17$\pm$5 & 0.51$\pm$0.22\\
K5\dotfill & 25$\pm$6 & 2.66$\pm$0.17 & 42$\pm$5 & 0.99$\pm$0.23\\
K3\dotfill & 29$\pm$8 & 4.16$\pm$0.5 & 65$\pm$6 & 1.4$\pm$0.5\\
\hline\noalign{\smallskip}
\multicolumn{5}{c}{1996.98.}\\
\hline\noalign{\smallskip}
Core\dotfill & 290$\pm$6 & 0.0 & 0.0 & $\la$0.01\\
K9\dotfill & 55$\pm$3 & 0.19$\pm$0.03 & 74$\pm$11 & $\la$0.01\\
K8\dotfill & 201$\pm$2 & 0.89$\pm$0.02 & 83$\pm$2 & 0.58$\pm$0.04\\
K7\dotfill & 132$\pm$2 & 0.91$\pm$0.03 & 54$\pm$2 & 0.84$\pm$0.07\\
K6\dotfill & 40$\pm$3 & 1.60$\pm$0.11 & 20$\pm$5 & 0.83$\pm$0.21\\
K5\dotfill & 20$\pm$6 & 2.47$\pm$0.35 & 46$\pm$11 & 0.97$\pm$0.75\\
K3\dotfill & 23$\pm$6 & 4.09$\pm$0.25 & 64$\pm$4 & 1.04$\pm$0.81\\
\noalign{\smallskip}
\hline
\end{tabular}
\end{flushleft}
\label{t22ghz}
\end{table}

\subsection{Changes in the jet trajectory}

The images of Figs. \ref{8ghz} and \ref{43ghz} confirm the twisted jet
structure previously observed in 0735+178 (G\'omez et al. 1999). A first
apparent bend of approximately 90$^{\circ}$ can be observed approximately 0.8
mas from the core, near the position of component K8 (k8 at 43 GHz), where the
jet direction changes from eastward to northward, toward component k7. This
curvature is more visible in the higher resolution 43 GHz images of
Fig. \ref{43ghz}. After this, the jet turns back toward the east, as traced by
several components identified at 8.4 GHz (see Fig. \ref{8ghz}).

Figure \ref{cp1} summarizes the components detected in 0735+178 since the
first observations in 1979. We have split the figure to cover two different
time spans. The top plot includes components observed between 1979.20 and
1992.44, while the bottom plot shows component positions after 1995.58 (no
observations have been published for epochs between 1992.44 and 1995.58). A
zoom showing more clearly the innermost components detected after 1995.58 is
plotted in Fig. \ref{cpin}.  Independent of the component identification
scheme, Fig. \ref{cp1} shows a qualitative change in the projected trajectory
of the jet in 0735+178. In the earlier epochs (top panel of Fig. \ref{cp1})
the jet components appeared to extend along a somewhat rectilinear trajectory
to the northeast, while after mid-1995 all observed components lie in a well
defined twisted structure, consistent with that mapped in our Figs. \ref{8ghz}
and \ref{43ghz}, resembling a helix in projection.

\subsection{Component identification}

We checked to see if any features previously identified in the VLBI jet of
0735+178 were still present in our new images by extrapolating the proper
motions detected for components K2 to K6 from previous epochs (B{\aa}{\aa}th
\& Zhang 1991; Zhang \& B{\aa}{\aa}th 1991; B{\aa}{\aa}th, Zhang, \& Chu 1991;
Gabuzda et al. 1994; G\'omez et al. 1999) to our new epochs. This procedure is
the basis for the possible component identifications shown in Figs. \ref{8ghz}
and \ref{43ghz} (see also Tables \ref{t8ghz}, \ref{t43ghz} and \ref{t22ghz}).
If this identification is correct, K2, K3, K4, K5, and K6 traveled downstream
with apparent velocities between $\sim$5-11 $h_{65}^{-1}$ c. Components K1,
K8, and k9 remained stationary within our position error estimates.

At the same time, Figs. \ref{cp1} and \ref{cpin} clearly show clustering of
components at several locations, labelled in Fig. \ref{cpin}. Thus, the data
after the middle of 1995 are also consistent with the VLBI jet of 0735+178
displaying primarily features that are stationary within the estimated
position errors.  In this case, components K4, K5, K6 and K6.5 of epoch 24 May
1998 would correspond to components K3, K4, K5 and K6 of the two previous
epochs, as indicated in Table~\ref{t8ghz} and Fig. \ref{cpin}. This
stationarity of components would contrast with the previously derived
superluminal motions, suggesting a possible change in the physical regime for
the VLBI jet flow.

It seems very likely that the presence of stationary components at the
locations of the two 90$^{\circ}$ bends (near K8 and K6 Fig.~\ref{cpin}) is
directly related to the bends in some way. One possibility is that there is
differential Doppler boosting in these locations, if these bends are
associated with jet regions where the flow velocity vectors are more closely
aligned with the line of sight to the observer (e.g. G\'omez et al. 1994;
Alberdi et al. 2000). Such bends in a highly relativistic flow would be
expected to lead to the formation of shocks, which could contribute to enhance
the emission at these locations. It is also possible that these bends
correspond to jet/external medium interactions, in which case strong shocks
are also expected. The stationarity of the remaining components might best be
interpreted in terms of standing shock waves. Numerical simulations show that
stationary recollimation shocks can be produced by pressure mismatches between
the jet and the external medium (G\'omez et al. 1995, 1997), as well as by jet
instabilities produced by the pass of strong plane perpendicular moving shocks
(Agudo et al. 2001). These recollimation shocks result in enhanced jet
emisson, which could be observed as stationary jet features.

Our new data thus open the possibility that there were mainly stationary
features in the VLBI jet of 0735+178 beginning sometime after the middle of
1992. This may suggest a change in the jet flow regime sometime after
mid-1992, since the clustering of components that is clearly visible in the
bottom panel of Fig.~\ref{cp1} and in Fig.~\ref{cpin} for epochs after
mid-1995 is not evident at earlier epochs. Comparing the models for the most
recent VLBI data, the simplest interpretation of the VLBI structure is
probably that the jet is made up of a series of stationary or quasi-stationary
components. At the same time, a plausible component identification scheme in
which the superluminal motions occurring at earlier epochs are continued at
the later epochs can be constructed, as discussed above. Clearly, further data
are needed to distinguish between these two possibilities.

\subsection{Stationary components}

\begin{figure*}
\centering
\mbox{\epsfxsize=5.6in\epsfbox[0 0 711 536]{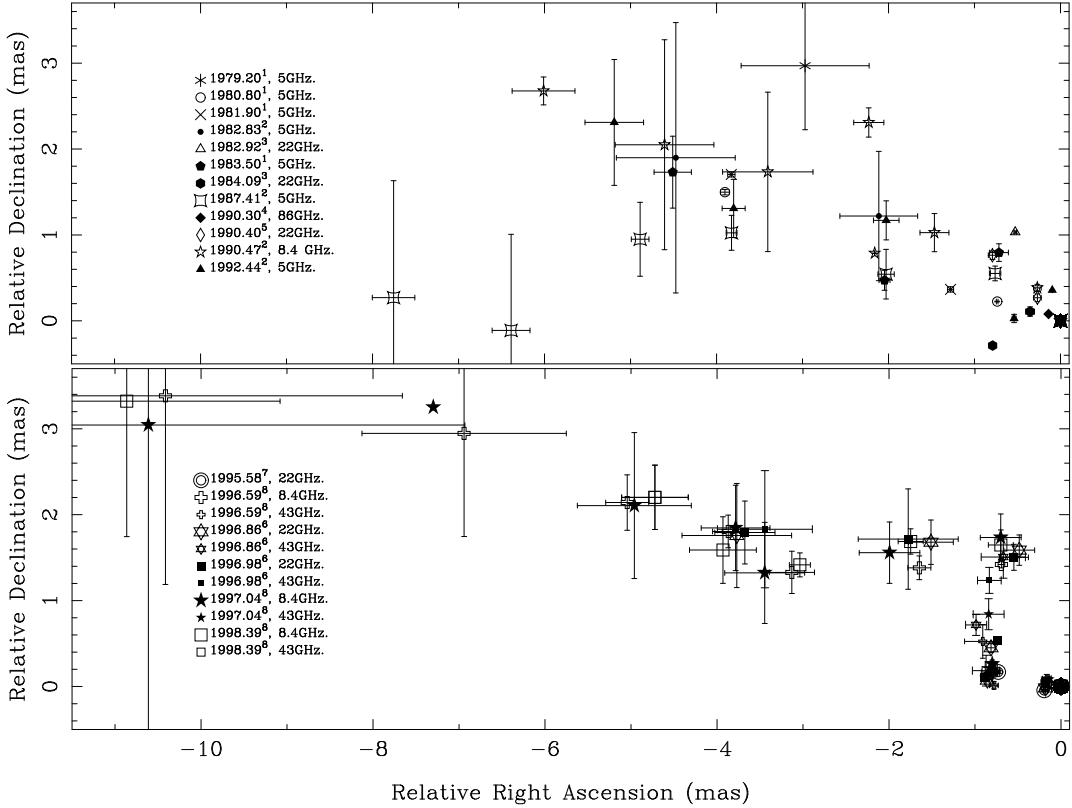}}
\caption{Components position, relative to the core, in the jet of 0735+178
detected between 1979.20 and 1992.44 ({\it top}), and between 1995.58 and
1998.39 ({\it bottom}). Epoch, frequency, reference, and symbol for each epoch
are as labeled. Estimated errors are indicated by bars. References correspond
to: 1. B{\aa}{\aa}th \& Zhang 1991; 2. Gabuzda et al. 1994; 3. Zhang \&
B{\aa}{\aa}th 1991; 4. Rantakyr\"o et al. 1998; 5. B{\aa}{\aa}th, Zhang, \&
Chu 1991; 6. G\'omez et al. 1999; 7. Gabuzda \& Cawthorne 2000; 8. This
paper.}
\label{cp1}
\end{figure*}

\begin{figure*}
\centering
\mbox{\epsfxsize=5.6in\epsfbox[0 0 494 576]{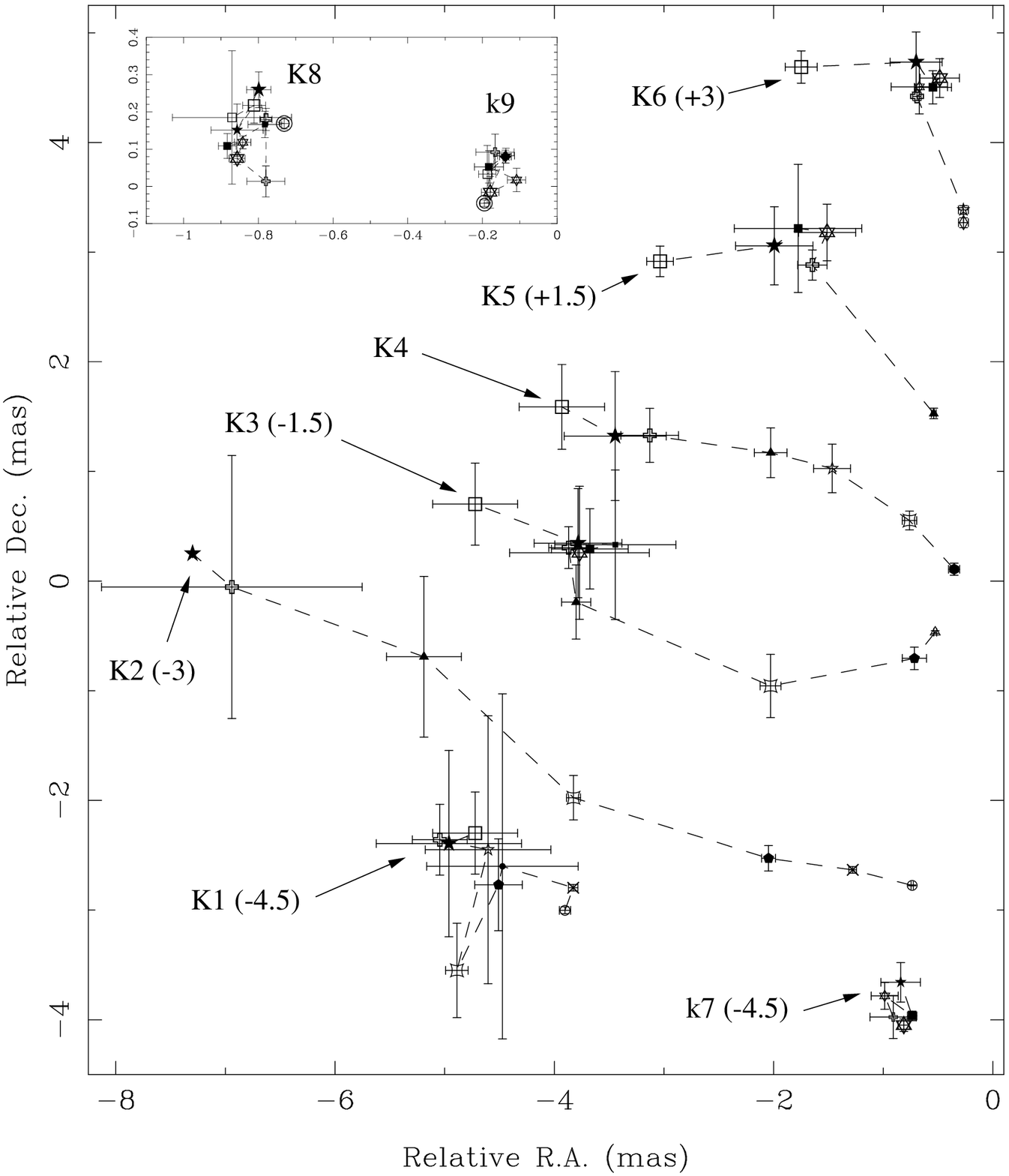}}
\caption{Apparent trajectories of components in 0735+178. Symbols used are the
same of Fig. \ref{cp1}. For clarity, components positions have been shifted
in declination by the amount (in milliarcseconds) expressed in
parentheses. Inset panels shows the fitted positions for components K8 (k8 at
43 GHz) and k9. Dashed lines connect the positions of the different components
across epochs.}
\label{cptrj}
\end{figure*}

Independent of the component identification scheme that is adopted, several
components---K1, K8/k8 and k9---have remained stationary within the errors. K1
was initially detected by B{\aa}{\aa}th \& Zhang \cite{BZ91} at four epochs
(their component B).  Their first identification for this component lay
significantly north of subsequent detections of K1 (see Fig. \ref{cp1}), which
would have required a very large apparent motion perpendicular to the observed
jet. We therefore suggest that the first identification for component K1 of
B{\aa}{\aa}th \& Zhang \cite{BZ91} may have corresponded to a different jet
feature, and we have not included it in our plot for K1 in Fig. \ref{cptrj}.

\begin{figure*}
\centering
\mbox{\epsfxsize=5.6in\epsfbox[0 0 722 373]{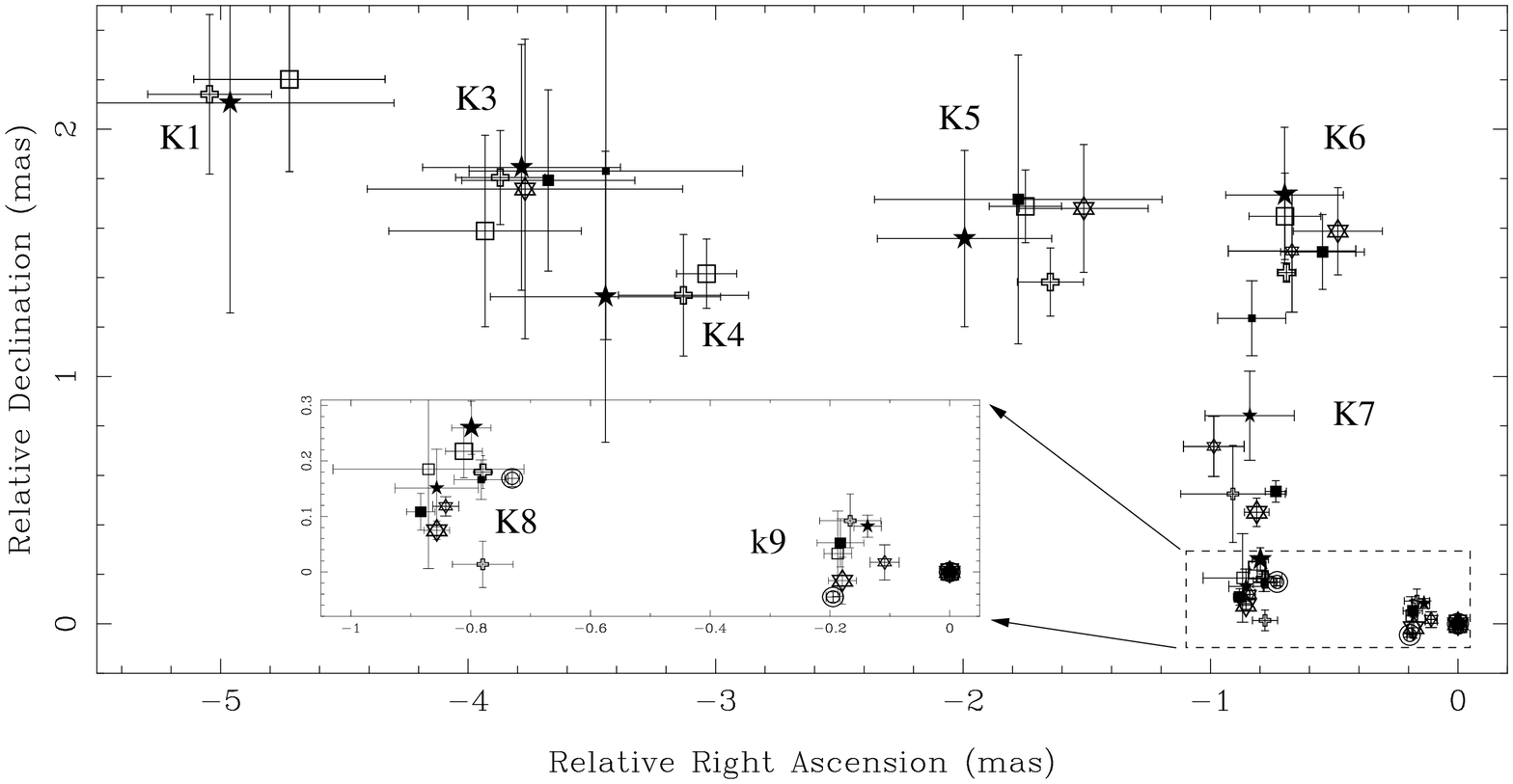}}
\caption{Zoom of the inner components detected between 1995.58 and 1998.39
(bottom panel in Fig. \ref{cp1}). Symbols are as in Fig. \ref{cp1}, and
labels correspond to the stationary-component scenario.}
\label{cpin}
\end{figure*}

Despite the stationarity of K1, in 1982.83 this component presented a small
downstream shift in position, remaining at the new location afterwards, within
the errors. The early flux density evolution for K1 (Gabuzda et al. 1994)
shows a rapid decrease from 1979.20 to 1980.8, followed by a slower decay
until it reaches a 5-GHz flux density of 80 mJy in 1983.5, which is roughly
mantained thereafter. Observations at 8.4 GHz since 1990.3 show small
variations in the flux density, which always remains below 40 mJy. 

Component K8/k8 was first detected in the 1995.58 22-GHz observations of
Gabuzda \& Cawthorne (2000, their component K1). The sequence of 15 and 22-GHz
images from 1996 presented by Homan et al. \cite{Ho01} shows a component
located at a mean distance of 0.8 mas in structural position angle
79.6$\pm$0.7$^{\circ}$, which could also be identified with K8/k8. Although
these authors measured a proper motion for this component of 0.14$\pm$0.03
mas/yr during 1996, a deceleration is observed in the last epochs. Subsequent
observations at 8.4, 22 and 43 GHz confirm the presence of this nearly
stationary component (see Figs. \ref{cpin} and \ref{cptrj}). K8 has an
optically thin spectrum between 8.4 and 43 GHz (see Tables~\ref{t8ghz} and
\ref{t43ghz}), although observations at 22 and 43 GHz show a slightly inverted
spectrum between these frequencies, which we find puzzling. Figure \ref{43ghz}
reveals that k8 is extended (see also Tables \ref{t8ghz}, \ref{t43ghz}, and
\ref{t22ghz}). It therefore seems likely that the ``wandering'' of its
position should be interpreted not as physical motion, but rather as the
result of internal changes in the component's brightness distribution, which
lead to small shifts in the position of the brightness centroid as a function
of frequency and epoch.

The innermost stationary component in 0735+178, k9, is $\sim$0.2 mas from the
core. Only observations above 22 GHz provide the necessary resolution to
detect it. It has an inverted spectrum and, like K8, shows small variable
offsets in position with epoch and frequency, which we also interpret as the
effect of changes in its internal brightness distribution. Observations at 86
GHz by Rantakyr\"o et al. \cite{Ra98} in 1990.3 revealed two strong components
separated by 0.16 mas. Although those observations and the observations of
Gabuzda \& Cawthorne {\cite{De00} are more than five years apart, it is
possible that component A of Rantakyr\"o et al. \cite{Ra98} should be
identified with component K2 of Gabuzda \& Cawthorne {\cite{De00} and our
component k9, suggesting that it has remained stationary for at least eight
years.  This component has a quasi-steady flux density and inverted spectrum.

Images at 22 and 43 GHz show a ``bridge'' of emission extending north to
connect component K8/k8 with the second 90$^{\circ}$ bend located near the
position of component K6 at epoch 1996.59. We have model fitted this emission
and identified it across epochs as k7. If component motions in 0735+178 are
non-ballistic, various components should have moved through the two apparent
90$^{\circ}$ bends at the positions of K8/k8 and K6, through where the
emission modelled as k7 is observed. However, k7 exhibits erratic motions
around the extended region, as indicated by the errors in its position. This
may suggest that what we have modelled as k7 actually corresponds to
underlying jet emission rather than a distinct feature in the flow.  This
region has a steep spectrum between 22 and 43 GHz.

The 8.4 GHz images in Fig. \ref{8ghz} show extended jet emission (with flux
density below $\sim$150 mJy) to the east of K2. To account for this flux
density, we have fitted its emission as a single component, labeled E, which
has a rather large flux density. Due to its very extended structure, the
errors in the model fitting are quite large. We emphasize that this is only
intended to allow for the presence of the extended emission in this region in
a general sense.

\subsection{Possible Superluminal components}

As discussed above, probably the simplest interpretation of the VLBI structure
observed at relatively recent epochs is that it is made up of a series of
stationary or quasi-stationary components. At the same time, we can plausibly
identify components across epochs in a way that implies superluminal motions
similar to those estimated for earlier epochs.  In this case (corresponding to
the component labels in Fig. \ref{8ghz}), components K2, K3, K4, K5 and K6 all
exhibit superluminal motion. As shown in Fig. \ref{cptrj}, all of these
components seem to follow curved paths, especially K5 and K6.

When estimating a component's implied trajectory, we averaged the positions 
obtained at different frequencies and nearby epochs. In this way, we  
derived mean positions for components detected at 1990.40 and 1990.47, 
and for 1996.86, 1996.98, and 1997.04. Typically, components expand as 
they move from the core, resulting in larger errors in the model fitting. 
To account for such variations in the position errors, we obtained a
weighted mean value for the apparent motions using the inverse square errors
as weights.  When computing the apparent speed, we allowed for motion
along non-rectilinear paths, and calculated the distance traveled in the 
plane of the sky. 

In this ``moving-component'' scenario, the first detection of K2 corresponds
to observations at 5 GHz by B{\aa}{\aa}th \& Zhang (1991, their component
C0). Observations at 5 GHz by Gabuzda et al. \cite{De94} in 1992.44 showed a
component at a distance of 5.68 mas from the core that they tentatively
identified with component K1. However, comparison with later epochs suggests
that this component may fit the trajectory and flux density evolution of K2
better, and therefore we consider this identification more plausible. Overall,
the trajectory for K2 is consistent with quasi-ballistic motion along position
angle 70$^{\circ}$. At epoch 1997.04, we could only obtain estimates of the
flux density and position of extended jet emission at the position expected
for K2. Due to the large errors associated with this component at this epoch,
no errors bars have been plotted in Figs. \ref{cp1} and \ref{cptrj}. The light
curve for component K2 shows a smooth decay with time. The analysis described
above yields an apparent speed for K2 of $11.6\pm 0.6 h_{65}^{-1}c$.

Overall, K3 follows a similar path to that of K2, but with a slower mean
apparent velocity of $8\pm 1.5 h_{65}^{-1} c$. Our last epoch suggests that K3
reached the position of K1, as reported previously for K2 by Gabuzda et al.
\cite{De94}. As in the K2+K1 intersection observed by Gabuzda et
al. \cite{De94}, our images do not show evidence of a violent interaction
between components K3 and K1. Table \ref{t8ghz} suggests that the flux density
of the merged K3 and K1 is approximately that of the sum of the two separate
components at epoch 1997.04. As shown in Fig. \ref{cptrj}, K1 remained
stationary (within the errors) during its intersection with K3. We note that
the inferred motion for K3 fluctuates between 1992.4 and 1998.4; this may
providing indirect evidence in favour of our alternative scenario, in which
the VLBI structure observed in recent epochs is made up of stationary or
quasi-stationary components.

Component K4 was first detected by Zhang \& B{\aa}{\aa}th \cite{ZB91} in
1984.09 (their component 2), 0.37~mas from the core. As shown in
Fig. \ref{cptrj}, the inferred trajectory is curved, and is consistent with
that outlined by the components in Figs. \ref{cp1} and \ref{cpin}.  However,
more frequent monitoring during its initial evolution would have been
necessary to test for the presence of non-ballistic motion through the inner
bends. The inferred flux density of K4 shows a monotonic decrease with time,
as expected for adiabatic evolution. This component's mean apparent speed is
$5\pm 1 h_{65}^{-1}c$.

Figure \ref{cptrj} shows that the moving-component scenario suggests very
non-ballistic trajectories for K5 and K6; however, this is based on an
uncertain identification with components at previous epochs. A time gap of
about 4.5 years separates the observations of Gabuzda et al. \cite{De94} and
of G\'omez et al. (1999; also Table \ref{t22ghz}). Our estimated apparent
speeds of $10\pm 3.5$ and $4.6\pm 0.1 h_{65}^{-1}c$ for K5 and K6,
respectively, are accordingly severely affected by this uncertainty in the
identifications with features at the earlier epochs. Component K6 was
initially identified as K5 at epoch 1990.47 by Gabuzda et al. \cite{De94}
based on its separation from the core. However, if we take into account its
structural position angle for that epoch, this component fits better with the
overall evolution of K6, and we have therefore identified it accordingly.

When analysing our images together with those published previously, we tried
to make our component identifications as consistent as possible with those
proposed by other authors. At the same time, due to the large time gaps
between some of the observing epochs, as well as the possible change in the
jet geometry that seems to have occurred in one of these gaps, there are
several components detected in earlier studies that do not fit easily the
behaviour inferred for our identified components. Those include component B of
B{\aa}{\aa}th \& Zhang \cite{BZ91}; component 3 of Zhang \& B{\aa}{\aa}th
\cite{ZB91}; component C2 of B{\aa}{\aa}th, Zhang, \& Chu \cite{Bal91}; and
components K2 (1982.83), C2 (1990.47), K3 (1990.47), K2.5 (1990.47), and K6
(1992.44) of Gabuzda et al. \cite{De94}. Some of these are from rather early
VLBI observations, and it may not be surprising if the corresponding images
and model fits were not entirely accurate representations of the true source
structure. In our ``moving components'' scenario, K6.5 is not identified with
any components seen in our or other studies, but it is identified with the
quasi--stationary component K6 in our ``stationary components'' scenario.

\section{Discussion}

Although it is quite common to observe relatively strong curvature in the jets
of compact AGN, the very twisted geometry found in 0735+178 presents a
somewhat dramatic and peculiar example, due to the presence of two
$90^{\circ}$ bends in the inner two mas of the VLBI jet. In addition,
comparison with previous observations suggests that the jet geometry has
changed with time. Figure \ref{cp1} reveals two distinct jet geometries before
and after the middle of 1992.  The observations of Gabuzda \& Cawthorne
\cite{De00} for 1995.58 and of Homan et al. \cite{Ho01} for several epochs in
1996 are consistent with the twisted inner jet geometry seen in our later
images, although they show only the inner 1 mas of the jet structure. The
5-GHz VLBA observations of Nan Rendong and collaborators at epoch 1995.33
(private communication) also confirm the presence of the two 90$^{\circ}$
bends, suggesting that if the jet in 0735+178 did experience a change in flow
regime, this occurred sometime between mid-1992 and mid-1995.

It seems likely that the apparent bending of the relativistic jet is so abrupt
(two bends through roughly $90^{\circ}$) because we are viewing the jet of
0735+178 at a small angle to the line of sight, so that intrinsic curvature is
enhanced by projection effects. The intrinsic bending could be produced either
by a change in the direction of ejection (e.g., jet precession or by other
more erractic variations) or by gradients in the external pressure that are
not aligned with the initial direction of the jet flow. Precession of the jet
could lead to either ballistic (as observed in SS~433) or non-ballistic fluid
motions, in which the flow velocity vector would follow the jet
bends. Three-dimensional numerical simulations of precessing jets (Aloy et
al. 1999, 2000) can give rise to helical structures, as produced by normal
mode jet instabilities, and have been applied to interpret the variability
observed in several sources (Hardee 2000). Non-ballistic motions would be
expected in this case.

Is the jet of 0735+178 precessing? Jet precession should lead to changes in
the inner jet geometry, as well as to components being ejected in different
position angles at different times, if the time scale for the precession is
comparable to that covered by the observations. Our observations suggest a
change in the apparent jet geometry in 0735+178, but the initial
quasi-rectilinear geometry observed in relatively early images does not appear
to be consistent with a precessing jet scenario. There is no clear and
consistent evidence that different components have been ejected in
systematically different directions with time, as would be expected for a
precessing jet. In addition, we find evidence for a number of stationary
components; if our identification of component k9 with component A of
Rantakyr\"o et al.  \cite{Ra98} is correct, this component remained stationary
for more than eight years.  A jet precessing in such a way as to lead to the
observed twisted inner-jet structure would require that k9 experience
significant changes in its position angle with respect to the core, as well as
flux variability, which are not observed. For this reason, it is most likely
that the jet of 0735+178 is not precessing, but additional observations are
needed to more conclusively test this hypothesis.

Are components in the jet of 0735+178 ballistic? Although earlier model fits
were consistent with roughly ballistic motion for a number of components, our
analysis of our more recent data indicates that either (1) the components in
the VLBI jet of 0735+178 have been roughly stationary at least since mid-1995
or (2) the VLBI jet components move superluminally along curved paths. The
possible co-existence of standing and moving features is more consistent with
non-ballistic component motions; the stationary components could be associated
with bends in the jet, as in the case of the quasi-stationary components K8
(at the first 90$^{\circ}$ bend, where the jet turns toward the north) and K6
(at the second 90$^{\circ}$ bend, where the jet turns back toward the east).
In addition, the jet of 0735+178 has stationary components in regions where
there is no apparent curvature. The proximity to the core of the stationary
component k9 suggests that it may be produced by a recollimation shock rather
than a bend. Relativistic numerical simulations (Agudo et al. 2001) suggest
that this type of stationary component, produced by jet instabilities, should
be common.

Overall, the stability of the jet structure over the many epochs of
observations made since mid-1995 (Fig.~\ref{cpin}) suggests that these images
show a series of quasi-stationary components, making the possibility that we
are actually viewing a superposition of linear ballistic trajectories that
``conspire'' to produce the observed structure unlikely. The polarimetric
observations of G\'omez et al. \cite{JL99} showed a longitudinal magnetic
field in the jet, which appeared to follow the curvature observed near K6,
also supporting a picture with non-ballistic fluid motions.

Since precession of the jet seems unlikely, it is more plausible that the
twisted geometry of 0735+178 is the result of pressure gradients in the
external medium through which the jet propagates, possibly triggered by its
own interaction with the ambient medium. In this case, we might expect gradual
changes in the position and curvature of the jet bends near K8 and K6 with
time. Evidence for the existence of such pressure gradients on VLBI scales is
provided, for example, by the jet/external medium interactions observed in
the radio galaxy 3C~120 (G\'omez et al.  2000), the compact steep-spectrum
source 3C~119 (Nan et al. 1999), the quasar 1055+018 (Attridge et al. 1999),
and the BL~Lac object 0820+225 (Gabuzda et al. 2001), and also the detection
of non-uniform parsec-scale rotation measures in a growing number of active
galactic nuclei (Nan et al. 1999; Taylor 1998, 2000; Zavala \& Taylor 2001;
Gabuzda et al. 2001; Reynolds et al. 2001).

The most intriguing result revealed by data obtained since mid-1995 is the
appearance of distinct clustering of components near several jet
locations. Although it is possible to derive a component identification scheme
for these recent images in which components move superluminally, this
clustering of components suggests instead a transition to a jet flow regime
that has led to the formation of a series of quasi-stationary regions of
emission. The clustering is rather stable, and it is unlikely that it could
arise by chance, due to particular epochs ``catching'' components right at
certain specific jet locations.  This component clustering opens a new view of
0735+178, in which essentially all jet components remain nearly stationary
with time, or at least have much smaller proper motions than measured
previously. This would require a re--evaluation of the physical parameters
estimated for 0735+178 from measuring proper motions, such as the observing
viewing angle, plasma bulk Lorentz factor, and those deduced from these.

The reason for the absence of this clustering of jet components in earlier
images remains unclear. It probably cannot be excluded that the quality of the
earlier images was such that the sharply twisted jet structure with
quasi-stationary regions of emission was present but remained
undetected. However, the second sharp bend is located well to the north of the
core, and it seems unlikely that it would not have been detected in any of the
earlier images, especially since some of the pre-1993 images were made at 22
GHz.  It seems more likely that the jet of 0735+178 experienced a change in
flow regime sometime between mid-1992 and mid-1995. If so, this is the first
time that such a transition has been observed. Continued monitoring of the
VLBI structure of 0735+178 is clearly of interest, both in order to obtain
further information about the origin of the twisted jet structure visible in
our images, and to see if the jet may make another transition in flow regime
that leads to the break-up of this stable twisted structure.

\section*{Acknowledgments}
This research was supported in part by Spain's Direcci\'on General de
Investigaci\'on Cient\'{\i}fica y T\'ecnica (DGICYT) grants PB97-1164 and
PB96-0782, by U.S. National Science Foundation grant AST-9802941, and by the
Fulbright commission for colaboration between Spain and the United States. DCG
acknowledges support from the European Commission under the IHP Programme
(ARI) contract No. HPRI-CT-1999-00045.

\end{document}